\documentclass{aa}  

\usepackage{graphicx}
\usepackage{chemformula}       % chemical formula
\usepackage{adjustbox}
\usepackage{listings}
\usepackage{xcolor} % So the code is coloured

\usepackage{txfonts}
\usepackage{threeparttable}
\usepackage{longtable}
\usepackage{booktabs}
\usepackage{ltablex}
\usepackage{caption}
\usepackage{hyperref}
\renewcommand\makeLineNumber{}

\begin{document} 

   \title{High-energy interstellar isomers: {\it cis-N-}methylformamide in the G+0.693-0.027 molecular cloud}

%   \subtitle{I. Overviewing the $\kappa$-mechanism}

\titlerunning{Detection of {\it cis-N-}methylformamide in G+0.693-0.027}

\authorrunning{Zeng et al.}

   \author{S. Zeng,\inst{1}
           V. M. Rivilla, \inst{2}
           M. Sanz-Novo, \inst{2}
           M. Melosso, \inst{3}
           I. Jim\'enez-Serra, \inst{2}
           L. Colzi, \inst{2}
           %J. Mart\'in-Pintado, \inst{2}
           A. Meg\'ias, \inst{2}
           D. San Andr\'es, \inst{2,4}
           A. L\'opez-Gallifa, \inst{2}
           A. Mart\'inez-Hernares, \inst{2}
           %B. Tercero, \inst{5}
           %P. de Vicente, \inst{6}
            \and
           S. Mart\'in. \inst{5,6}
           %M. A. Requena-Torres\inst{9,10}
          }

        \institute{Star and Planet Formation Laboratory, Pioneering Research Institute (PRI), RIKEN, 2-1 Hirosawa, Wako, Saitama, 351-0198, Japan\\
              \email{shaoshan.zeng@riken.jp}
         \and
        Centro de Astrobiolog\'ia (CAB), INTA-CSIC, Carretera de Ajalvir km 4, Torrej\'on de Ardoz, 28850, Madrid, Spain
         \and
        Dipartimento di Chimica ``Giacomo Ciamician'', Universit\`a di Bologna, via P. Gobetti 85, 40129 Bologna, Italy
         \and
        Departamento de F{\'i}sica de la Tierra y Astrof{\'i}sica, Facultad de Ciencias F{\'i}sicas, Universidad Complutense de Madrid, 28040 Madrid, Spain
         \and
        European Southern Observatory, Alonso de C\'ordova 3107, Vitacura 763 0355, Santiago, Chile
         \and
        Joint ALMA Observatory, Alonso de C\'ordova 3107, Vitacura 763 0355, Santiago, Chile
        \\
             }

   %\date{Received September 15, 1996; accepted March 16, 1997}

% \abstract{}{}{}{}{} 
% 5 {} token are mandatory
 
  \abstract
   {Isomerism in complex organic molecules provides key insights into the formation mechanisms and physical conditions of the interstellar medium (ISM). Among the C$_2$H$_5$NO isomers, only acetamide and \textit{trans}-N-methylformamide (\textit{trans}-NMF) have been detected in space. The recent detection of higher-energy isomers in other chemical families raises questions about the formation and abundance of less stable isomers.}
   {We aim to search for {\it cis-N-}methylformamide (\textit{cis}-NMF), the next higher-energy conformer in the C$_2$H$_5$NO family and investigate its possible formation pathways.}
   {We used ultra-sensitive wide-band spectral surveys obtained with the Yebes 40 m and IRAM 30 m telescopes to search for \textit{cis-}NMF towards the Galactic Centre molecular cloud G+0.693-0.027. A spectroscopic catalogue was extrapolated from literature data to aid the search.}
   {We present the first detection of \textit{cis-}NMF in the ISM, with 55 unblended or slightly blended transitions, 44 of which were new transitions identified based on extrapolated spectroscopic data. Due to the lack of collisional rate coefficients, a quasi-non-LTE analysis, which separated the transitions into different \textit{K}$_a$ ladders, was used to determine the excitation conditions. The derived column density is (1.5$\pm$0.1) $\times$ 10$^{13}$ cm$^{-2}$, corresponding to a molecular abundance of (1.1$\pm$0.2) $\times$ 10$^{-10}$ relative to H$_2$. The resulting \textit{trans/cis-}NMF isomeric ratio of 2.9$\pm$0.6 deviates significantly from thermodynamic expectations, suggesting that kinetic non-equilibrium processes and stereospecific chemical pathways are responsible for the formation of \textit{cis-}NMF in this environment.}
   {The detection of \textit{cis-}NMF expands the known inventory of interstellar C$_2$H$_5$NO isomers and challenges the assumption that isomer abundances strictly correlate with thermodynamic stability. Laboratory and theoretical studies propose formation via CH$_3$NCO hydrogenation or spin-forbidden reactions involving CH$_2$ and NH$_2$CHO, though these may not reflect typical ISM conditions. This finding highlights the need for further investigation into isomerisation mechanisms and constrains astrochemical models of complex organic molecules.}
   {}

   \keywords{ISM: molecules - Astrochemistry - Molecular data - 
               }

   \maketitle
%
%-------------------------------------------------------------------
\section{Introduction} \label{sec:intro}

Isomerism refers to molecules that share the identical molecular formula, they contain the same number and types of atoms, but have different arrangements. In general, there are two main types of isomers: constitutional (or structural) isomers and stereoisomers (or spatial isomers). Structural isomers differ in connectivity, meaning they contain the same components but are bonded together in different ways. As a result, they are typically considered distinct molecules with different physical and chemical properties. Spatial isomers, on the other hand, have the exact same connectivity but differ in the spatial orientation of their constituting atoms and hence are expected to exhibit more similarities than structural isomers. A specific type of stereoisomer, spatial isomers (e.g., $E$/$Z$ and \textit{cis}/\textit{trans} isomers), have substituent groups fixed in particular positions. In \textit{cis}/\textit{trans} isomerism, the groups are positioned on the same side (\textit{cis}) or opposite sides (\textit{trans}) of a reference plane. Alternatively, the \textit{E/Z} notation is used to describe the relative positions of high-priority groups based on the Cahn-Ingold-Prelog priority rules \citep{10.1039/9781849733069}. 

From an astronomical perspective, one important application of isomers is their use as a valuable tool for directly probing the physical and chemical processes occurring in the regions where they reside within the interstellar medium (ISM). In particular, since isomers share a similar level of chemical complexity and consist of the same constituent atoms, their conversion barriers and zero-point energy differences can primarily predict their existence and help elucidate the formation mechanisms responsible for the observed abundance ratios. By investigating the relative abundance ratio of 32 isomers across 14 species, \citet{Lattelais2009,Lattelais2010} proposed the concept of the `minimum energy principle' (hereafter MEP), which is based on two main concepts: (1) the most thermodynamically stable isomer, i.e., with the lowest zero-point energy ($\Delta E$), should be the most abundant; and (2) the relative abundance ratio between the most stable isomer and its higher-energy isomers appears to be closely correlated with the energy differences between them. Whilst the MEP has been shown to be generally applicable across various astronomical environments, studies of several families of structural isomers have cast doubt on this principle. Not only have the less stable, high-energy isomers, those predicted to be less likely observed, been robustly detected, but similar abundances have also been found between these high-energy isomers and their more stable counterparts \citep[e.g., isomers of C$_3$H$_2$O, C$_2$H$_4$O$_2$, and C$_2$H$_5$NO$_2$;][]{Loomis2015, Bermudez2018, Mininni2020, Cabezas2021, Shingledecker2020, Rivilla2023}.

In the case of spatial isomers, their isomeric abundance ratio, such as the $\textit{anti}$ and $\textit{gauche}$ conformers of ethyl formate \citep[C$_2$H$_5$OCHO)][]{Rivilla2017}, the $\textit{Aa}$ and $\textit{Ga}$ conformers of $\textit{n-}$propanol \citep[$\textit{n-}$C$_3$H$_7$OH][]{Jimenez-Serra2022}, the $\textit{cis}$ and $\textit{trans}$ conformers of thioformic acid \citep[HC(O)SH][]{Garcia2022}, and the $\textit{Z}$ and $\textit{E}$ isomers of imines \citep{Garcia2021,San-Andres2024} are plausibly explained by their predicted relative stabilities. In these cases, thermodynamic equilibrium can be reached especially when ground-state quantum tunnelling effects are taken into account \citep{Garcia2021,Garcia2022}. However, an increasing number of exceptions suggest that, rather than following thermodynamics, the observed isomeric abundance ratios may result from selective competitive chemical pathways \citep[e.g., the $\textit{cis}$ and $\textit{trans}$ isomers of methyl formate (HC(O)OCH$_3$), formic acid (HCOOH), and the $\textit{cis}$-$\textit{cis}$ and $\textit{cis}$-$\textit{trans}$ isomers of carbonic acid (HOCOOH)][]{Neill2012, Garcia2022, Garcia2023, Sanz-Novo2025}. Altogether, these findings highlight a new opportunity to detect higher-energy isomers in the ISM. Detecting such isomers, especially those expected to be in low abundance or unlikely to be detected based on the MEP, will help further elucidate the key factors that govern isomeric abundance ratios in interstellar environments, whether thermodynamic stability is adhered or stereospecific chemical pathways must be considered. 

Motivated by the recent detection of the higher-energy trans-isomer of methyl formate \citep[$\Delta E$ = 25 kJ$\,$mol$^{-1}$, or $\sim$3000 K;][]{Neill2012} as well as the identification of several amides, including acetamide (CH$_3$CONH$_2$) and \textit{trans-}NMF, in the Galactic Centre molecular cloud G+0.693-0.027 (hereafter G+0.693) \citep{Zeng2023,Sanz-Novo2025b}, we report the first detection of \textit{cis-}N-methylformamide (\textit{cis-}CH$_3$NHCHO; hereafter \textit{cis-}NMF) in the ISM. G+0.693 is a chemically rich molecular cloud located within the Sgr B2 star-forming complex in the Central Molecular Zone (CMZ), the inner $\sim$600 pc of the Galaxy. Over the past decade, numerous molecular species, including more than 25 newly identified in the ISM, have been detected towards this source \citep[e.g.][and references therein]{Jimenez-Serra2022,Rivilla2023,Zeng2023,San-Andres2024,Sanz-Novo2025b}.

\textit{cis-}NMF belongs to the C$_2$H$_5$NO isomeric family and is the third member of this group to be detected in the ISM to date \citep{Lattelais2010,Foo2018}. The most stable isomer, acetamide, along with \textit{trans-}NMF, lying 10.1 kcal$\,$mol$^{-1}$ ($\sim$5100 K) higher in energy than acetamide, have previously been detected in high-mass star-forming regions \citep{Hollis2006,Halfen2011,Belloche2017,Ligterink2020,Colzi2021}, in intermediate-mass regions \citep{Ligterink2020}, and in G+0.693 itself \citep{Zeng2023}. \textit{cis-}NMF lies $\sim$1.33 kcal$\,$mol$^{-1}$ ($\sim$670 K) above \textit{trans-}NMF \citep{Kawashima2010}, and is both a structural isomer of acetamide and a spatial isomer of \textit{trans-}NMF. The detection of \textit{cis-}NMF in the same physical environment as its more stable counterparts enables a direct comparison of excitation conditions and molecular abundances across the isomeric series, offering new constraints on the formation pathways and chemical stability of prebiotic molecules in the ISM.

%-----------------------------------------------------------------
\begin{figure}
\centering
\includegraphics[width=0.8\linewidth]{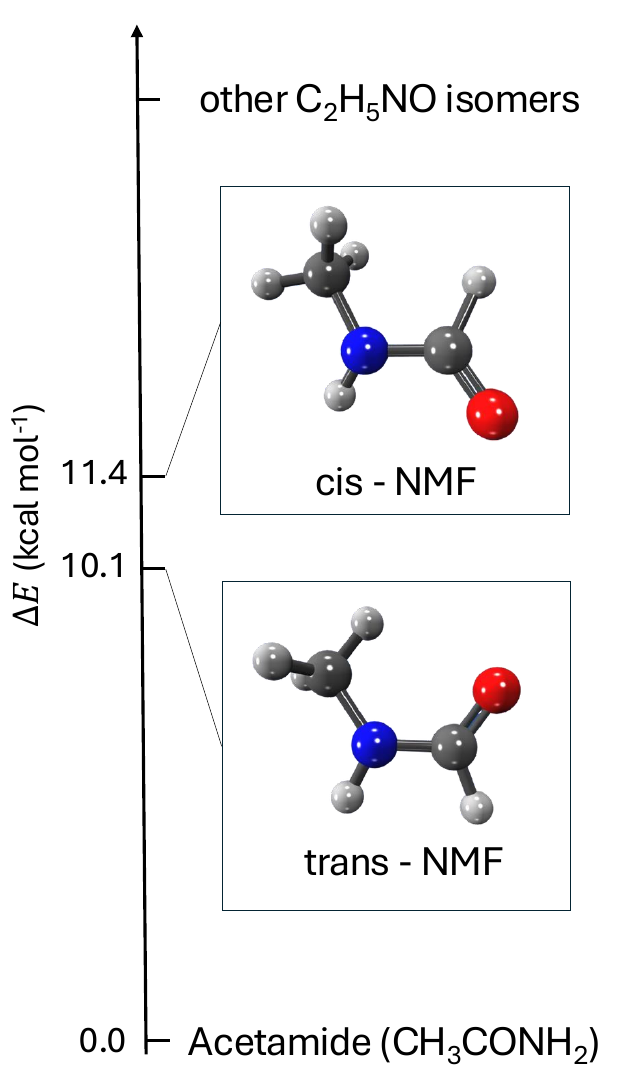}
\caption{The structure of \textit{cis-}NMF (top) and \textit{trans-}NMF (bottom), optimized at the B2PLYPD3/aug-cc-pVTZ level of theory, with nitrogen atom in blue, oxygen atom in red, hydrogen atoms in white, and carbon atoms in grey.}
\label{fig:trans_cis_NMF}
\end{figure}
%-----------------------------------------------------------------

\section{Observations} \label{sec:obs}
We used the unbiased spectral surveys performed with Yebes 40$\,$m\footnote{The 40$\,$m radiotelescope at Yebes Observatory is operated by the Spanish Geographic Institute (IGN, Ministerio de Transportes, Movilidad y Agenda Urbana.) \url{http://rt40m.oan.es/rt40m en.php}} and IRAM 30$\,$m\footnote{IRAM is supported by INSU/CNRS (France), MPG (Germany), and IGN (Spain)} telescopes to search for the \textit{cis} conformer of CH$_3$NHCHO towards G+0.693 molecular cloud. The observations were centred at $\alpha$(J2000) = 17$\rm^h$47$\rm^m$22$\rm^s$, $\delta$(J2000) = -28$\rm^{\circ}$21$\rm^{\prime}$27$\rm^{\prime\prime}$ and the position switching mode was employed with the reference position of $\Delta\alpha$, $\Delta\delta$ = $-$885$^{\prime\prime}$, 290$^{\prime\prime}$ with respect to the source position. The line intensity of the spectra was measured in units of antenna temperature (\textit{T}$_{\rm A}^*$) since the molecular emission towards G+0.693 is extended over the beam \citep{Jones2012,Li2020,Zeng2020,Zheng2024}. 

The Yebes 40m observations (project 21A014; PI Rivilla) were performed in multiple sessions between March 2021 and March 2022. The Nanocosmos Q-band (7 mm) HEMT receiver was employed to provide ultra broad-band observations (18.5 GHz) in two linear polarisations \citep{Tercero2021}. The Fast Fourier Transform Spectrometers (FFTS) backends were used, providing a raw channel width of 38 kHz. The observations have a total frequency coverage of 31.07 $-$ 50.42 GHz and the half power beam width (HPBW) of the telescope was $\sim$35$^{\prime\prime} - $55$^{\prime\prime}$ across the observed frequency range. The final spectra were smoothed to 256 kHz which is equivalent to velocity resolutions of 1.5 $-$ 2.5 km s$^{-1}$ and the achieved root-mean-square (rms) noise level lies between 0.25 $-$ 0.9 mK depending on the frequency.

For the IRAM 30m observations (project 123-22; PI: Jim\'enez-Serra), the Eight MIxer Receiver (EMIR) was connected to the Fast Fourier Transform Spectrometer (FTS200) to provide a channel width of 200 kHz. In this work, data covering the spectral windows from 71.8 to 116.7 GHz were used and the corresponding HPBW of telescope was $\sim$21$^{\prime\prime} - $34$^{\prime\prime}$. The spectra were smoothed to velocity resolutions of 1.0 $-$ 2.6 km s$^{-1}$ and the achieved rms noise level lies between 1.3 $-$ 2.8 mK depending on the frequency. 

%%%%%%%%%%%%%%%%%%%%%%%%%%%%%%%
\begin{figure*}
\includegraphics[width=\textwidth]{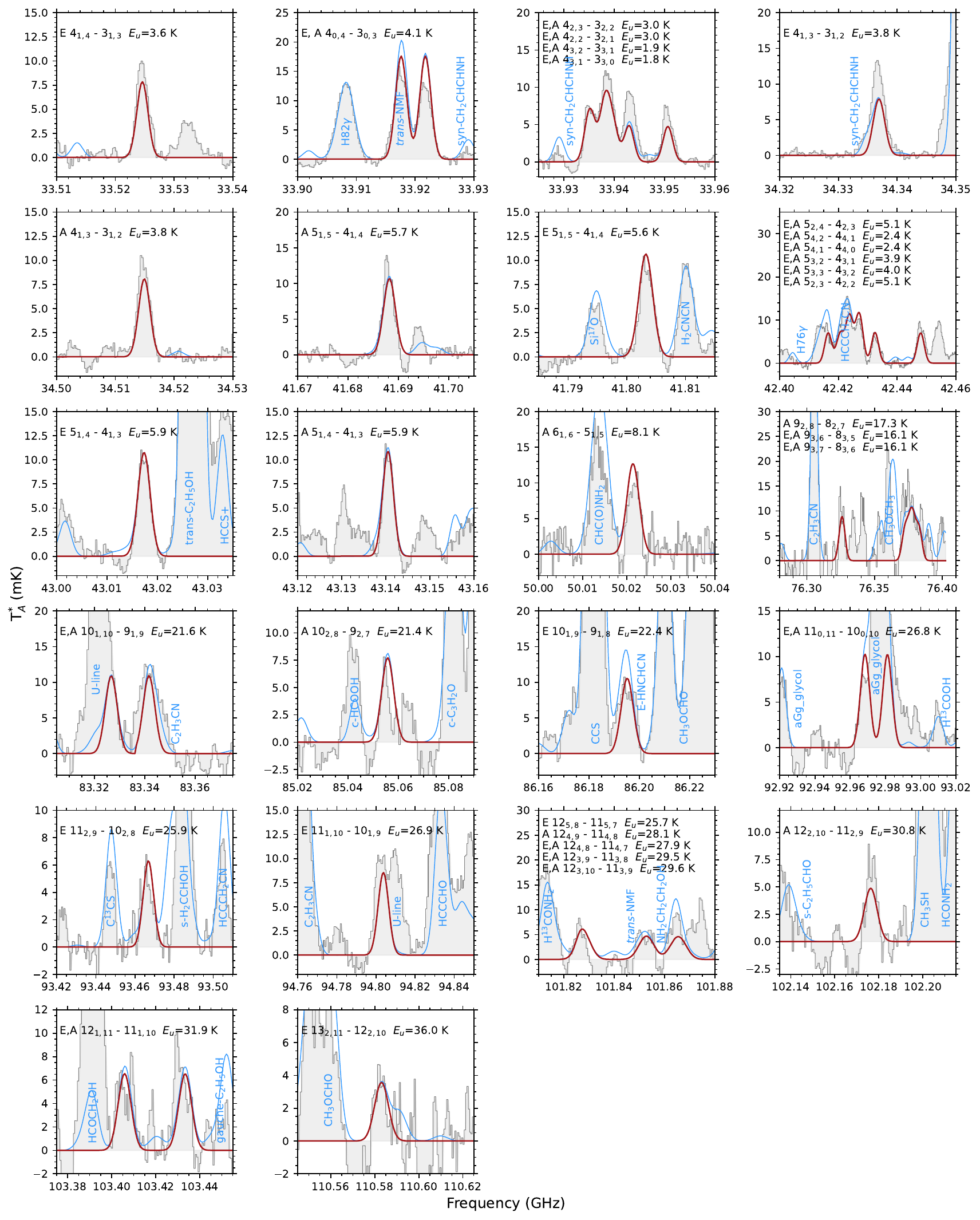}
\caption{Unblended or slightly blended transitions of \textit{cis-N-}methylformamide detected towards G+0.693 arranged in order of increasing frequency. The grey histogram and shaded area correspond to the observed spectra. The blue lines depict the synthetic spectra, accounting for contributions from all detected molecular species (over 140) in G+0.693. The red line indicates the best LTE fit provided by \textsc{madcuba}. The quantum numbers and $E_u$ values for each detected transition are listed in the upper left corner of each panel.
\label{fig:line}}
\end{figure*}
%%%%%%%%%%%%%%%%%%%%%%%%%%%%%%%

\section{Analysis and Results} 
\label{sec:results}

\subsection{Spectroscopic data of \textit{cis-}NMF}
\label{subsec:spec}
The rotational spectrum of \textit{cis}-NMF was recorded between 8 and 35 GHz by Fourier transform microwave spectroscopy \citep{Kawashima2010}. About one hundred transitions, exhibiting resolved nitrogen hyperfine structure and methyl internal rotation splitting, were assigned and successfully analysed using the rho-axis method. With the aim of preparing a spectral line catalogue to be used for astronomical purposes, we have re-fitted the transitions observed by \citet{Kawashima2010} using the \texttt{XIAM} code \citep{Hartwig1996}, which employs the internal axes method (IAM, \citealt{Woods1966, Vacherand1986}) instead. Our results are equivalent in terms of quality of the fit, the rms error of the lines being around 18 kHz. The spectroscopic parameters determined with \texttt{XIAM} and the electric dipole moment components computed in \citet[][$\mu_a$ = 4.45 D and $\mu_b$ = 0.45 D at the MP2/6-31G$^{**}$ level of theory]{Kawashima2010} have subsequently been used to predict the rotational spectrum of \textit{cis}-NMF in the 31–116 GHz range. Since molecular lines are typically observed towards G+0.693 with a full width at half maximum (FWHM) around 20 km s$^{-1}$, the effect of the hyperfine structure has been ignored as it cannot be spectrally resolved. Finally, the spectral predictions have been reformatted in order to match the structure of a standard *.cat file \citep{Pickett1991}.

To refine the $\Delta E$ between the two NMF conformers, we have also performed new electronic structure computations using the software package Gaussian 16 \citep{g16}. Particularly, we have conducted single-point energy calculations at the CCSD(T)/aug-cc-pVTZ level of theory on top of the B2PLYPD3/aug-cc-pVTZ optimized geometries (shown in Figure \ref{fig:trans_cis_NMF}). We have also computed harmonic vibrational frequencies at the B2PLYPD3/aug-cc-pVTZ level of theory to account for zero-point vibrational energy (ZPE) corrections. Thus, we obtained a $\Delta E$ = 1.4 kcal$\,$mol$^{-1}$, which is in close agreement with the previously reported value of 1.3 kcal$\,$mol$^{-1}$ computed by \cite{Kawashima2010} at the MP2/6-31G$^{**}$ level of theory.

\subsection{Detection of \textit{cis-}NMF}
\label{subsec:det}
To search for \textit{cis-}NMF, the prepared spectroscopic catalogue was imported into the Spectral Line Identification and Modelling (SLIM) tool within \textsc{Madcuba} package \footnote{Madrid Data Cube Analysis on ImageJ is a software developed at the Center of Astrobiology (CAB) in Madrid; \url{http://cab.inta-csic.es/madcuba/}}\citep[version 15/06/2024,][]{Martin2019}. In Table \ref{tab:obs_transitions}, the unblended or slightly blended transitions of \textit{cis-}NMF detected towards G+0.693 are listed with relevant spectroscopic information. We adopted the same criteria for identifying unblended or slightly blended transitions as defined by \citet{Rey-Montejo2024} and \citet{Sanz-Novo2025}. Figure \ref{fig:line} shows, in the order of increasing frequency, the best fitted line profiles of \textit{cis-}NMF. The rest of the \textit{cis-}NMF transitions that are covered in the spectral range of the survey are consistent with the observed spectra, being either blended with brighter transitions from other molecules or too weak to be securely detected. To accurately evaluate the potential line contamination by other species, emission from over 140 molecular species that are already identified towards G+0.693 are considered in the overall fitting. 

In total, 55 (28 E-state and 27 A-state) unblended or slightly blended transitions of \textit{cis-}NMF are detected with an integrated signal-to-noise (S/N)\footnote{The S/N ratio is computed from the integrated intensity ($\int T^{*}_{A} {\rm d}\nu$) and noise level, $\sigma$ = rms $\times$ $\sqrt{\delta\nu \times \mathrm{FWHM}}$, where $\delta \nu$ is the spectral resolution and the FWHM is obtained from the best fitting.} ratio $\geq$5. Among the detected transitions, 11 below 33.95 GHz correspond to those previously measured in laboratory spectroscopic studies, while over 40 transitions at frequencies above 33.95 GHz have been identified in this work based on extrapolated spectroscopic data. These detections represent the first direct identification of such transitions in the ISM, significantly extending the known spectroscopic range of \textit{cis-}NMF.

Initially, all the line profiles were fitted under the assumption of Local Thermodynamic Equilibrium (LTE) conditions in order to derive the physical parameters, using the \textsc{autofit} tool within \textsc{madcuba-slim}, which performs nonlinear least-squares LTE fitting based on the Levenberg–Marquardt algorithm. However, several high \textit{K}$_a$ transitions were poorly fitted, suggesting that they are likely influenced by non-LTE effects. Since the collisional rate coefficients are not available, we performed a so-called quasi-non-LTE analysis which involve in splitting the transitions into different $K_a$ rotational ladders  (i.e. $K_a$ = 0, 1, 2, etc.). Similar approaches have already been applied for the analysis of other molecules towards this cloud \citep[e.g.][]{Zeng2018, Rodriguez-Almeida2021a, Sanz-Novo2025b} and appear as the only viable procedure to determine the excitation conditions of this molecule in G+0.693. A \textsc{Madcuba} script, based on some of the newly developed tools, is provided as supplementary information in Appendix \ref{app}. It enables the automatic separation of the transitions into the different $K_a$ ladders and re-computation of the rotational partition function accordingly. The derived physical parameters of \textit{cis-}NMF are listed in Table \ref{tab:raio}. 

To achieve the best LTE modelling, the radial velocity ($\nu_{\rm LSR}$) and the full width at half maximum (FWHM) were fixed at 68 km s$^{-1}$ and 20 km s$^{-1}$, respectively, while the excitation temperature ($T_{\rm ex}$) and column density ($N$) were left as free parameters. The resulting $T_{\rm ex}$ values derived from each individual $K_a$ ladder range from 8 to 14 K, which is consistent with those obtained for all amide species as well as other complex organic molecules (COMs) detected towards G+0.693 \citep[e.g.][]{Requena-Torres2008,Rivilla2022c,Zeng2023}. The total molecular column density, computed as the sum of the column density of all the individual $K_a$ ladders, is estimated to be (1.5$\pm$0.1) $\times$ 10$^{13}$ cm$^{-2}$. This translates to a molecular abundance with respect to H$_2$ of (1.1$\pm$0.2) $\times$ 10$^{-10}$, assuming N$\rm _{H_2}$ = 1.35 $\times$ 10$^{23}$ cm$^{-2}$ with an uncertainty of 15$\%$ \citep{Martin2008}. Adopting the molecular abundance of \textit{trans-}NMF = (3.2$\pm$0.3) $\times$ 10$^{-10}$ from \citet{Zeng2023}, the \textit{trans/cis-}NMF ratio is derived to be 2.9$\pm$0.6.

\section{Discussion} \label{subsec:dis}

Due to the lack of available spectroscopic data, higher-energy isomers of the C$_2$H$_5$NO family, other than acetamide, \textit{trans}-NMF, and \textit{cis}-NMF, cannot yet be searched for. Table \ref{tab:raio} summarises their relative stabilities in terms of $\Delta E$, dipole moment ($\mu$), derived column densities, and the corresponding abundances relative to H$_2$ towards G+0.693. According to the MEP, the most stable isomer should be the most abundant. In this regard, the observed molecular abundance of these three C$_2$H$_5$NO isomers appear to follow the principle qualitatively, with lower abundances corresponding to higher energies. However, if the isomeric abundance ratio is assumed to be governed by thermodynamic equilibrium, it should scale with exp(-$\Delta E/T_{k}$), where $\Delta E$ is the zero-point energy difference and $T_{k}$ is the gas kinetic temperature. Considering that T$_{k}$ of G+0.693 is in the range of 70-140 K \citep{Zeng2018} and $\Delta E$(\textit{trans/cis}) = 670 K, the thermodynamically expected \textit{trans/cis-}NMF ratio should should range from 120 to 14350, more than an order of magnitude higher than the actual observed value of 2.9$\pm$0.6. In addition, the direct isomerisation of \textit{trans-}NMF has an energy barrier of $\sim$20 kcal\,mol$^{-1}$ (or 10300$\,$K) \citep{Tsai2022}, which is unlikely to be overcome under typical ISM conditions. It is thus implicitly suggested that thermodynamic equilibrium cannot be attained between \textit{cis/trans-}NMF. 

\textit{cis}-NMF has largely been overlooked in both observational and theoretical studies, primarily due to its higher energy and presumed low abundance, which rendered it unlikely to be detectable in the ISM. As a result, current chemical models \citep[e.g.][]{Belloche2019, Garrod2022} have typically included only the \textit{trans} conformer, which was the sole isomer detected prior to this study. Consequently, \textit{trans}-NMF has generally been treated as the representative NMF species in astrochemical modelling. Although no efficient gas-phase formation route is currently known for NMF, several formation pathways on grain surfaces have been proposed. These include radical–radical reactions such as HNCHO + CH$_3$ \citep{Belloche2017, Garrod2022} and CH$_3$NH + HCO \citep{Frigge2018}. HNCHO may form either through radical addition (e.g. NH + CHO) or via cosmic ray-induced photodissociation of NH$_2$CHO, while CH$_3$NH is thought to result from the irradiation of CH$_3$NH$_2$ by energetic electrons. Another proposed grain-surface pathway involves the hydrogenation of CH$_3$NCO \citep{Belloche2017}. A more detailed discussion of these formation mechanisms in the context of G+0.693 can be found in \citet{Zeng2023}.

One available study addressing the formation of \textit{cis}-NMF is the laboratory experiment conducted by \citet{Tsai2022}, which demonstrated a hydrogen-atom-assisted isomerisation mechanism. The proposed scheme begins with barrierless H-abstraction of \textit{trans-}NMF, leading to the formation of the \textit{trans-}\ch{.C(O)NHCH3} radical, followed by a second H-abstraction to produce CH$_3$NCO. Once CH$_3$NCO is formed, hydrogenation results in the \textit{cis-}\ch{.C(O)NHCH3} radical, which subsequently undergoes another hydrogenation to yield \textit{cis-}NMF. Although the experiment was conducted in solid para-hydrogen—an environment not directly representative of astronomical conditions—it provides an important insight into the formation mechanism of NMF isomers. Specifically, due to steric hindrance, the hydrogenation of CH$_3$NCO was found to yield exclusively \textit{cis-}NMF, without producing the \textit{trans-}conformer. This result is in agreement with laboratory ice experiments, where CH$_3$NCO is efficiently formed but \textit{trans-}NMF remains undetected in CH$_4$:HNCO ice mixtures, suggesting that hydrogenation of CH$_3$NCO in interstellar ice analogues may not lead to the formation of \textit{trans-}NMF \citep{Ligterink2018}. 

In G+0.693, CH$_3$NCO has been detected with a relative abundance of 4.9$\times$10$^{-10}$ \citep{Zeng2018}, which is about one order of magnitude higher than that of \textit{cis-}NMF and less than a factor of two higher than \textit{trans-}. However, this abundance pattern alone is insufficient to confirm the formation pathway. Further laboratory studies involving the hydrogenation of pure CH$_3$NCO under astrophysical relevant conditions are required to determine whether this process can account for the observed abundances of both NMF conformers.

Another study, based on computational investigation, proposed that a spin-forbidden reaction between CH$_2$ and NH$_2$CHO could serve as a potential formation mechanism for both \textit{trans-}NMF and \textit{cis-}NMF in the gas phase \citep{Mirzanejad2025}. According to the calculations, the overall formation pathway is barrierless. It first produces an intermediate that exists as two chiral isomers, each of which can proceed to a transition state structure, ultimately forming the non-chiral \textit{trans-}NMF and \textit{cis-}NMF via hydrogen abstraction. The calculations suggest that the formation of \textit{cis-}NMF is more favourable, as its pathway is barrierless, whereas the formation of \textit{trans-}NMF involves a barrier of 1.9 kcal$\,$mol$^{-1}$ (or 956 K) and thus depends on the probability of hydrogen tunnelling through this barrier. However, as noted by the authors, \textit{trans-}NMF is $\sim$1.33 kcal$\,$mol$^{-1}$ more stable than \textit{cis-}NMF. As a result, the rate constants for the formation of both isomers are expected to be comparable, suggesting that they could form in similar abundances. However, the detection of both \textit{trans-}NMF and \textit{cis-}NMF in G+0.693, with observed abundances differing by a factor of three, indicates that this formation pathway alone cannot fully account for the observations. Moreover, as the proposed mechanism involves a spin-forbidden reaction, a relatively uncommon process in astrochemical models, further detailed kinetic studies are essential. In particular, investigations into spin-inversion probabilities, reaction rate constants, and quantum tunnelling efficiencies are needed to better understand the observed abundance ratio and to provide deeper insight into the formation mechanisms of these isomers in the ISM.

%%%%%%%%%%%%%%%%%%%%%%%%%%%%%%%
\begin{table*}
    \centering
      \caption{Relative stabilities, dipole moments, and abundances of detected C$_2$H$_5$NO isomers towards G+0.693. \label{tab:raio}}
    \begin{adjustbox}{width=\textwidth}
     \begin{tabular}{cccccccc}
\hline     
    Molecule & $\mu_a,\mu_b$$^{a}$ & $\Delta E$$^{b}$ & \textit{T}$\rm _{ex}$ & $\nu \rm_{LSR}$ & FWHM & $\textit{N}$$^{d}$ &  $\textit{X}$\\
    & (Debye) & (kcal mol$^{-1}$) & (K) & (km s$^{-1}$) & (km s$^{-1}$) & $\times$10$^{12}$ (cm$^{-2}$) & $\times$10$^{-10}$ \\
    \hline
     CH$_3$C(O)NH$_2$ & 1.22, 3.47 & 0.0 & 7.2-7.9 & 68.4-69.0 & 18.9-22.0 & 115$\pm$2 & 8.5$\pm$1.3 \\
     %\vspace{5pt} \\
     \textit{trans-}NMF & 3.2, 2.4 & 10.1 & 7.1$\pm$0.4 & 68.2$\pm$0.5 & 19$\pm$1 & 43$\pm$4 & 3.2$\pm$0.6 \\
     %\vspace{5pt} \\
     \textit{cis-}NMF (total) & 4.3 (4.45), 0.2 (0.45) & 11.4 & - & - & - & 14.5$\pm$0.7 & 1.1$\pm$0.2\\
     \textit{cis-}NMF (\textit{K}$_a$=0) & - & - & 8.5$\pm$0.8 & 68$^{c}$ & 20$^{c}$ & 4.2$\pm$0.6 & -\\
     \textit{cis-}NMF (\textit{K}$_a$=1) & - & - & 11.0$\pm$0.6 & 68$^{c}$ & 20$^{c}$ & 4.1$\pm$0.2 & -\\
     \textit{cis-}NMF (\textit{K}$_a$=2) & - & - & 11.0$\pm$0.5 & 68$^{c}$ & 20$^{c}$ & 2.8$\pm$0.1 & -\\
     \textit{cis-}NMF (\textit{K}$_a$=3) & - & - & 8.6$\pm$0.6 & 68$^{c}$ & 20$^{c}$ & 2.3$\pm$0.2 & -\\
     \textit{cis-}NMF (\textit{K}$_a$=4) & - & - & 14$\pm$3 & 68$^{c}$ & 20$^{c}$ & 0.55$\pm$0.12 & -\\
     \textit{cis-}NMF (\textit{K}$_a$=5) & - & - & 14$^{c}$ & 68$^{c}$ & 20$^{c}$ & 0.53$\pm$0.09 & -\\
     \hline\hline
    \end{tabular}
    \end{adjustbox}
        \begin{minipage}{\textwidth}
    \footnotesize
    $^{a}$ Values for CH$_3$C(O)NH$_2$ are taken from \citet{Kojima1987}, while those for \textit{trans-} and \textit{cis-}NMF are based on B2PLYPD3/aug-cc-pVTZ calculations performed in this work. Values in parentheses correspond to those reported by \citet{Kawashima2010}. $^{b}$ Values taken from \citet{Lattelais2010},\citet{Kawashima2010}, and this work. $^{c}$ Value fixed in the \textsc{madcuba} fit. $^{d}$ \textit{N}$\rm _{H_2}$ = 1.35 $\times$ 10$^{23}$ cm$^{-2}$ with an uncertainty of 15$\%$ is adopted from \citet{Martin2008}. 
        \end{minipage}
\end{table*}
%%%%%%%%%%%%%%%%%%%%%%%%%%%%%%%

\section{Conclusions} \label{subsec:con}
We report the first interstellar detection of cis-N-methylformamide (\textit{cis-}NMF, \textit{cis-}CH$_3$NHCHO) towards the Galactic centre molecular cloud G+0.693-0.027, marking it as the third confirmed isomer of the C$_2$H$_5$NO family observed in space. Using ultra-sensitive spectral data from the Yebes 40 m and IRAM 30 m telescopes, we identify 55 unblended or slightly blended transitions of \textit{cis-}NMF and determine a column density of (1.5$\pm$0.1) $\times$ 10$^{13}$ cm$^{-2}$. This corresponds to an abundance of (1.1$\pm$0.2) $\times$ 10$^{-10}$ relative to H$_2$, resulting in a \textit{trans/cis} abundance ratio of 2.9$\pm$0.6.

It is worth noting that existing spectroscopic studies have only reported transitions up to 33.95 GHz. Consequently, any transitions detected in this work beyond that frequency represent the first direct identifications in the ISM. Although these detections are based on extrapolated spectroscopic data, the broad linewidth typically observed towards G+0.693 mitigate any significant impact on our analysis. However, this highlights a critical need for the laboratory spectroscopic study to extend measurements into the millimetre-wave regime. Expanding the available spectroscopic datasets will be essential for improving the accuracy and reliability of molecular identifications in future astronomical surveys, particularly in chemically rich environments like G+0.693.

The resulting \textit{trans/cis} ratio of 2.9$\pm$0.6 determined in G+0.693 is qualitatively consistent with MEP, yet it deviates significantly from the value expected under thermodynamic equilibrium. This discrepancy suggests that the observed abundance pattern cannot be attributed solely to thermal stability. Instead, the data support a scenario in which non-equilibrium chemical processes and stereospecific reaction pathways play a major role in shaping the isomeric distribution. The detection of a higher-energy isomer such as \textit{cis}-NMF further challenges once again the general applicability of the MEP in the ISM and reinforces the need for more comprehensive kinetic modelling.

Although laboratory and theoretical studies suggest that formation routes involving hydrogen-atom-assisted isomerisation or barrierless radical reactions could preferentially lead to the production of \textit{cis}-NMF, the former does not reflect an astronomically relevant environment, while the latter involves a relatively unconventional mechanism. More broadly, the limited detections of molecules, particularly higher-energy isomers, pose significant constraints on the investigation of their formation pathways. The chemistry of not only \textit{trans}/\textit{cis}-NMF but also other isomers of C$_2$H$_5$NO remains poorly understood. This discovery adds an important piece to the puzzle of complex organic chemistry in space and highlights the chemical richness of G+0.693 as a valuable natural laboratory for studying conformational isomerism and molecular evolution in the ISM.

\begin{acknowledgements}
We thank the anonymous referee for a careful review. S. Z. acknowledge the support by RIKEN Special Postdoctoral Researchers Program. V.M.R., L.C, I.J-.S, A.M., D.S.A, and A. L-G. acknowledge support from the grant PID2022-136814NB-I00 by the Spanish Ministry of Science, Innovation and Universities/State Agency of Research MICIU/AEI/10.13039/501100011033 and by ERDF, UE; V.M.R. D.S.A., and A. L-G. acknowledge the funds provided by the Consejo Superior de Investigaciones Cient{\'i}ficas (CSIC) and the Centro de Astrobiolog{\'i}a (CAB) through the project 20225AT015 (Proyectos intramurales especiales del CSIC). V.M.R. also acknowledges the grant RYC2020-029387-I funded by MICIU/AEI/10.13039/501100011033 and by "ESF, Investing in your future", and from the grant CNS2023-144464 funded by MICIU/AEI/10.13039/501100011033 and by “European Union NextGenerationEU/PRTR”. I.J-.S and A.M. also acknowledges support from ERC grant OPENS, GA No. 101125858, funded by the European Union. Views and opinions expressed are however those of the authors only and do not necessarily reflect those of the European Union or the European Research Council Executive Agency. D.S.A. also extends his gratitude for the financial support provided by the Comunidad de Madrid through the Grant PIPF-2022/TEC-25475. M. S.-N. also acknowledges a Juan de la Cierva Postdoctoral Fellow proyect JDC2022-048934-I, funded by MICIU/AEI/10.13039/501100011033 and by the European Union “NextGenerationEU/PRTR”. M.M. thanks the European Union -- Next Generation EU under the Italian National Recovery and Resilience Plan (PNRR M4C2, Investment 1.4 -- Call for tender n. 3138 dated 16/12/2021—CN00000013 National Centre for HPC, Big Data and Quantum Computing (HPC) -- CUP J33C22001170001). 
\end{acknowledgements}

\bibliography{references}{}
\bibliographystyle{aa}

\begin{appendix}
\label{app}

\section{List of unblended or slightly blended transitions of \textit{cis-}NMF detected towards G+0.693-0.027.}

Table \ref{tab:obs_transitions} presents the spectroscopic information on unblended or slightly blended transitions of \textit{cis-}NMF identified in G+0.693. It includes the rest frequencies of the transitions, quantum numbers, the base-10 logarithm of the integrated intensity at 300 K (log I), the energies of the upper levels for each transition (E$_u$), and any molecular species blended with the transitions.

%%%%%%%%%%%%%%%%%%%%%%%%%%%%%%%
\begin{table*}[htb!]
  \centering
  \caption{Spectroscopic information on the unblended or slightly blended transitions of \textit{cis-}NMF detected towards G+0.693.}
  \label{tab:obs_transitions}
  \begin{tabular}{rrrrcrrrccrc}
    \hline\hline \\
     \multicolumn{1}{c}{Frequency} & \multicolumn{8}{c}{Transition$^{a}$} & log $I$ & \multicolumn{1}{c}{$E_u$} & Blending \\
     \multicolumn{1}{c}{(GHz)} & $J^\prime$ & $K_a^\prime$ & $K_c^\prime$ & & $J$ & $K_a$ & $K_c$ & (A/E) & (nm$^2$MHz) & \multicolumn{1}{c}{(K)} & \\[1.0ex]
     \hline \\[-1.5ex]
     33.524476 &   4 & 1 &  4 & - &  3 & 1 &  3 & (E) &  -5.064  & 3.6  &  unblended \\        
     33.917500 &   4 & 0 &  4 & - &  3 & 0 &  3 & (E) &  -5.016  & 4.0  &  \textit{trans-}NMF \\  
     33.921622 &   4 & 0 &  4 & - &  3 & 0 &  3 & (A) &  -5.016  & 4.0  &  unblended \\        
     33.934793 &   4 & 2 &  3 & - &  3 & 2 &  3 & (A) &  -5.152  & 3.0  &  unblended \\        
     33.935534 &   4 & 3 &  1 & - &  3 & 3 &  0 & (E) &  -5.399  & 1.8  &  unblended \\    
     33.937748 &   4 & 3 &  2 & - &  3 & 3 &  1 & (A) &  -5.399  & 1.9  &  unblended \\    
     33.937780 &   4 & 3 &  1 & - &  3 & 3 &  0 & (A) &  -5.399  & 1.9  &  unblended \\    
     33.938784 &   4 & 2 &  2 & - &  3 & 2 &  1 & (E) &  -5.152  & 3.0  &  unblended \\    
     33.940034 &   4 & 3 &  2 & - &  3 & 3 &  1 & (E) &  -5.399  & 1.6  &  unblended \\    
     33.942841 &   4 & 2 &  3 & - &  3 & 2 &  2 & (E) &  -5.151  & 2.8  &  unblended \\    
     33.950541 &   4 & 2 &  2 & - &  3 & 2 &  1 & (A) &  -5.151  & 3.0  &  unblended \\    
     34.336790 &   4 & 1 &  3 & - &  3 & 1 &  2 & (E) &  -5.043  & 3.8  &  \textit{syn-}NH$_2$COCH$_2$OH\\
     34.514804 &   4 & 1 &  3 & - &  3 & 1 &  2 & (A) &  -5.032  & 3.8  &  unblended \\            
     41.688134 &   5 & 1 &  5 & - &  4 & 1 &  4 & (A) &  -4.763  & 5.6  &  unblended \\            
     41.803109 &   5 & 1 &  5 & - &  4 & 1 &  4 & (E) &  -4.764  & 5.6  &  unblended \\            
     42.416372 &   5 & 2 &  4 & - &  4 & 2 &  3 & (A) &  -4.814  & 5.0  &  H76$\gamma$\\        
     42.419774 &   5 & 4 &  2 & - &  4 & 4 &  1 & (A) &  -5.216  & 2.4  &  HCCCH$_2$CN\\        
     42.419775 &   5 & 4 &  1 & - &  4 & 4 &  0 & (A) &  -5.216  & 2.4  &  HCCCH$_2$CN\\        
     42.419814 &   5 & 4 &  1 & - &  4 & 4 &  0 & (E) &  -5.215  & 2.2  &  HCCCH$_2$CN\\        
     42.420734 &   5 & 3 &  2 & - &  4 & 3 &  1 & (E) &  -4.946  & 3.8  &  HCCCH$_2$CN\\        
     42.423474 &   5 & 3 &  3 & - &  4 & 3 &  2 & (A) &  -4.946  & 3.9  &  HCCCH$_2$CN\\        
     42.423586 &   5 & 3 &  2 & - &  4 & 3 &  1 & (A) &  -4.946  & 3.9  &  HCCCH$_2$CN\\        
     42.424596 &   5 & 4 &  2 & - &  4 & 4 &  1 & (E) &  -5.215  & 2.0  &  HCCCH$_2$CN\\        
     42.426391 &   5 & 3 &  3 & - &  4 & 3 &  2 & (E) &  -4.946  & 3.6  &  unblended \\        
     42.427326 &   5 & 2 &  3 & - &  4 & 2 &  2 & (E) &  -4.814  & 5.0  &  unblended \\        
     42.432313 &   5 & 2 &  4 & - &  4 & 2 &  3 & (E) &  -4.814  & 4.8  &  unblended \\        
     42.447860 &   5 & 2 &  3 & - &  4 & 2 &  2 & (A) &  -4.814  & 5.0  &  unblended \\        
     43.017211 &   5 & 1 &  4 & - &  4 & 1 &  3 & (E) &  -4.740  & 5.8  &  unblended \\        
     43.140362 &   5 & 1 &  4 & - &  4 & 1 &  3 & (A) &  -4.733  & 5.8  &  unblended \\        
     50.021233 &   6 & 1 &  6 & - &  5 & 1 &  5 & (A) &  -4.523  & 8.0  &  unblended \\        
    %76.158936 &   9 & 0 &  9 & - &  8 & 0 &  8 & (E) &  -3.98  & 14.6  &  \\        
    %76.168928 &   9 & 0 &  9 & - &  8 & 0 &  8 & (A) &  -3.98  & 14.6  &  \\        
     76.325738 &   9 & 2 &  8 & - &  8 & 2 &  7 & (A) &  -4.011  & 17.2  &  unblended \\        
     76.372075 &   9 & 3 &  6 & - &  8 & 3 &  5 & (E) &  -4.054  & 16.1  &  unblended \\        
     76.375974 &   9 & 3 &  7 & - &  8 & 3 &  6 & (A) &  -4.054  & 16.1  &  unblended \\        
     76.378422 &   9 & 3 &  6 & - &  8 & 3 &  5 & (A) &  -4.054  & 16.1  &  unblended \\        
     76.382617 &   9 & 3 &  7 & - &  8 & 3 &  6 & (E) &  -4.053  & 15.8  &  unblended \\        
     83.326407 &  10 & 1 & 10 & - &  9 & 1 &  9 & (A) &  -3.869  & 21.6  &  U-line \\        
     83.341379 &  10 & 1 & 10 & - &  9 & 1 &  9 & (E) &  -3.869  & 21.6  &  C$_2$H$_3$CN \\          
    %84.571143 &  10 & 0 & 10 & - &  9 & 0 &  9 & (E) &  -3.849  & 22.3  &  \\        
    %84.582463 &  10 & 0 & 10 & - &  9 & 0 &  9 & (A) &  -3.849  & 22.3  &  \\      
    %84.840589 &  10 & 5 &  6 & - &  9 & 5 &  5 & (A) &  -4.041  & 16.6  &  \\        
    %84.840589 &  10 & 5 &  5 & - &  9 & 5 &  4 & (A) &  -4.041  & 16.6  &  \\      
    %84.846323 &  10 & 5 &  5 & - &  9 & 5 &  4 & (E) &  -4.04  & 16.3  &  \\        
    %84.850127 &  10 & 4 &  7 & - &  9 & 4 &  6 & (A) &  -3.966  & 18.7  &  \\        
    %84.850153 &  10 & 4 &  6 & - &  9 & 4 &  5 & (A) &  -3.966  & 18.7  &  \\        
    %84.850185 &  10 & 4 &  6 & - &  9 & 4 &  5 & (E) &  -3.966  & 18.5  &  \\        
    %84.851799 &  10 & 5 &  6 & - &  9 & 5 &  5 & (E) &  -4.04  & 16.2  &  \\        
    %84.859958 &  10 & 4 &  7 & - &  9 & 4 &  6 & (E) &  -3.966  & 18.3  &  \\        
     85.055480 &  10 & 2 &  8 & - &  9 & 2 &  7 & (A) &  -3.873  & 21.3  &  unblended \\          
     86.194881 &  10 & 1 &  9 & - &  9 & 1 &  8 & (E) &  -3.840  & 22.4  &  $E$-HNCHCN \\        
     92.968105 &  11 & 0 & 11 & - & 10 & 0 & 10 & (E) &  -3.732  & 26.8  &  $aGg^\prime$-(CH$_2$OH)$_2$ \\        
     92.980811 &  11 & 0 & 11 & - & 10 & 0 & 10 & (A) &  -3.732  & 26.8  &  $aGg^\prime$-(CH$_2$OH)$_2$ \\        
    %93.326389 &  11 & 5 &  7 & - & 10 & 5 &  6 & (A) &  -3.898  & 21.1  &  \\        
    %93.326389 &  11 & 5 &  6 & - & 10 & 5 &  5 & (A) &  -3.898  & 21.1  &  \\        
     93.466530 &  11 & 2 &  9 & - & 10 & 2 &  8 & (E) &  -3.754  & 25.8  &  unblended \\          
     94.803456 &  11 & 1 & 10 & - & 10 & 1 &  9 & (E) &  -3.722  & 26.9  &  U-line \\        
   %101.362827 &  12 & 0 & 12 & - & 11 & 0 & 11 & (A) &  -3.626  & 31.6  &  \\      
    101.826220 &  12 & 5 &  8 & - & 11 & 5 &  7 & (E) &  -3.773  & 25.6  &  unblended \\    
    101.827581 &  12 & 4 &  9 & - & 11 & 4 &  8 & (A) &  -3.717  & 28.0  &  unblended \\      
    101.827658 &  12 & 4 &  8 & - & 11 & 4 &  7 & (E) &  -3.717  & 27.8  &  unblended \\      
    101.827678 &  12 & 4 &  8 & - & 11 & 4 &  7 & (A) &  -3.717  & 28.0  &  unblended \\      
    101.851638 &  12 & 3 &  9 & - & 11 & 3 &  8 & (E) &  -3.674  & 29.5  &  \textit{trans-}NMF \\      
    101.853270 &  12 & 3 & 10 & - & 11 & 3 &  9 & (A) &  -3.675  & 29.6  &  \textit{trans-}NMF \\      
    101.863947 &  12 & 3 &  9 & - & 11 & 3 &  8 & (A) &  -3.674  & 29.6  &  NH$_2$CH$_2$CH$_2$OH \\      
    101.866176 &  12 & 3 & 10 & - & 11 & 3 &  9 & (E) &  -3.674  & 29.3  &  NH$_2$CH$_2$CH$_2$OH \\      
   %102.013689 &  12 & 2 & 10 & - & 11 & 2 &  9 & (E) &  -3.645  & 30.7  &  \\      
    102.175979 &  12 & 2 & 10 & - & 11 & 2 &  9 & (A) &  -3.642  & 30.7  &  unblended \\      
    103.405488 &  12 & 1 & 11 & - & 11 & 1 & 10 & (E) &  -3.615  & 31.9  &  unblended \\      
    103.432992 &  12 & 1 & 11 & - & 11 & 1 & 10 & (A) &  -3.615  & 31.9  &  unblended \\      
    110.582703 &  13 & 2 & 11 & - & 12 & 2 & 10 & (E) &  -3.546  & 36.0  &  unblended \\
	\hline\hline \\[-3.5ex]
		\end{tabular}
		\tablefoot{$^{a}$ The rotational energy levels of each transition are given using the standard notation for asymmetric tops exhibiting internal rotation of a methyl group: $J$ is the quantum number denoting the total angular momentum; \textit{K}$_a$ and \textit{K}$_c$ are pseudo-quantum numbers needed for labelling purposes; A/E denotes the A and E torsional sublevels associated to the methyl internal rotation of \textit{cis-}NMF.}
\end{table*}
%%%%%%%%%%%%%%%%%%%%%%%%%%%%%%%

\section{\textsc{Madcuba} script for the \textit{K}$_a$ ladder separation analysis}

The detailed \textsc{Madcuba} script used in this study to perform the \textit{K}$_a$ ladder separation analysis for \textit{cis-}NMF. It first separated the \textit{K}$_a$ ladder automatically. For each \textit{K}$_a$ ladder, the script first corrects the  energy of the rotational levels by substracting the energy of the lower-state energy level  (\textit{E}$_L$), and then recalculates the partition function as a direct summation of all the energy levels within each ladder. The \textsc{autofit} tool was then applied to obtain the best-fitting LTE model for the transitions of each \textit{K}$_a$ ladder, and the total column density was derived as a sum of the column density of each ladder.

\begin{lstlisting}[caption={\textit{K}$_a$ ladder separation}, label={madcuba_script_1}]

# Running the macro in \textsc{madcuba}
call('MADCUBA_IJ.setActiveInformationMADCUBA',false);

# Input file
spectraFile='full_survey12may2023.fits'  

# Output file
pathbase="/Users/Kladderseparation/"; 

# Name and catalogue of the molecule
molecule = "USER$cis-N-CH3NHCHO";  

# Maximum value of \textit{K}$_a$ to be included in the analysis
k_max = "6";     

# To rename the molecule and separate the \textit{K}$_a$ = X transitions
var quote = "'"
new_mol=substring(molecule, indexOf(molecule, "$")+1);
molecules = quote+molecule+"$Any$Any$Any$"+quote; 
rename_molecule = quote+new_mol+"_all"+quote;  
f = File.open(pathbase + "/Kladder_separation_analysis.txt");

print("-------------------------");
print("----K ladder separation----");
print("-------------------------");
print(f,"------------------------");
print(f,"----K ladder separation----");
print(f,"------------------------");

for (i=0; i<k_max; i++) { // 0; i<
   rename_molecule = quote+new_mol+"_K"+i+quote;
   cond = quote+"CAST (id_QNN2 AS INTEGER)="+i+quote;   

 print("----"+rename_molecule+"----");
 print(f,"----"+rename_molecule+"----");

# Perform the SLIM search for the transitions belonging to each Ka ladder fall within the data. For found transitions, substract the Elow for each Ka ladder and update the partition function of the entry, which is computed by direct summation of the energy of all the levels included in each ladder

 run("SLIM Search", "range='selected_data' rename_molecule="+rename_molecule+" where="+cond+" axislabel='Frequency' axisunit='Hz' molecules="+molecules+"  searchtype=add datafile="+spectraFile+" datatype=SPECTRA");
 print(rename_molecule+ " is Search="+ call("SLIM_Search.existsResult"));

 existsSearch = call("SLIM_Search.existsResult");
 if(existsSearch=='true')
 {
  run ("SLIM Update Q Elow", "molecule="+rename_molecule+" catalog='USER'  subtract_elow=true where="+cond+" partition=total");
  print("E_LOW_K_ladder (cm-1) ="+call("SLIM_Update_Q_Elow.getSubtractElow"));
  print("E_LOW_K_ladder (K) ="+(parseFloat(call("SLIM_Update_Q_Elow.getSubtractElow"))*1.438777));
  print(f,"E_LOW_K_ladder (cm-1) ="+call("SLIM_Update_Q_Elow.getSubtractElow"));
  print(f,"E_LOW_K_ladder (K) ="+(parseFloat(call("SLIM_Update_Q_Elow.getSubtractElow"))*1.438777));
  }
}

print("-------------------------");
print("----End of K ladder separation----");
print("-------------------------");
print(f,"------------------------");
print(f,"----End of K ladder separation----");
print(f,"------------------------");

File.close(f); 

\end{lstlisting}

\begin{lstlisting}[caption={Compute molecular column density}, label={madcuba_script_2}]
# Define the name of the molecule, the path of the output, the total number of Ka ladders

molecule = "USER$cis-N-CH3NHCHO";
num_k = 6; 
pathbase="/Users/Kladderseparation/"; 

var quote = "'" 
new_mol=substring(molecule, indexOf(molecule, "$")+1);
molecules = quote+molecule+"$Any$Any$Any$"+quote; 

f = File.open(pathbase + "/Results.txt");

energy = newArray(num_k);
logNladder = newArray(num_k);
elogNladder = newArray(num_k);
numFit = 0;
qLogN = 0;
eqlogN = 0;

# Reconstrut the name of each entry and compute the total column density of the molecule as a sum of the column density of each Ka ladder

for (i = 0; i < num_k; i++) {
    rename_molecule = quote+new_mol+"_K"+i+quote;
    run("SLIM Select Molecule", "molecule=" + rename_molecule);

    if (call('SLIM_Parameters.getValue', 'Autofit') == "true") {
        logNladder[numFit] = log(10) * parseFloat(call('SLIM_Parameters.getValue', 'logN|EM'));
        elogNladder[numFit] = log(10) * parseFloat(call('SLIM_Parameters.getValue', 'delta logN|EM'));
        energy[numFit] = 1.438777 * parseFloat(call('SLIM_Parameters.getValue', 'Eladder'));
        qLogN += pow(10, parseFloat(call('SLIM_Parameters.getValue', 'logN|EM')));
        eqLogN += pow(10, parseFloat(call('SLIM_Parameters.getValue', 'delta logN|EM')));
    numFit++;
    }
}

# Adjust the arrays to the number of Ka ladders considered and print everything in the .log and in the output file

energy= Array.trim(energy,numFit);
logNladder= Array.trim(logNladder,numFit);
elogNladder= Array.trim(elogNladder,numFit);
//Array.show("DR values ",energy , logNladder,elogNladder,cqlogNladder,ecqlogNladder);

print("-------------------------");
print("----Results----");
print("-------------------------");
print("Number of K ladders included in the analysis: " + num_k);
print("-------------------------");
print("LogNtot (cm-2) : " + log(qLogN) / log(10));
print("Ntot (cm-2): " + qLogN);
print("errorNtot (cm-2): " + eqLogN);


print("-------------------------");
print("----Termination of Results----");
print("-------------------------");

print(f,"--------------------------");
print(f,"----Results----");
print(f,"--------------------------");
print(f,"Number of K ladders included in the analysis: " + num_k);
print(f,"--------------------------");
print(f,"LogNtot (cm-2) : " + log(qLogN) / log(10));
print(f,"Ntot (cm-2): " + qLogN);
print(f,"errorNtot (cm-2): " + eqLogN);

print(f,"--------------------------");
print(f,"----Termination of Results----");
print(f,"--------------------------");

File.close(f); 

\end{lstlisting}

\end{appendix}

\end{document}